\documentclass[journal]{IEEEtran}
\usepackage{graphicx}
\usepackage{amsmath}
\usepackage{amsfonts}
\usepackage{multirow}
\usepackage{booktabs}  

\begin{document}

\title{Adaptive Progressive Attention Graph Neural Network for EEG Emotion Recognition}
\author{Tianzhi Feng, Chennan Wu, Yi~Niu, Fu Li$^*$, Yang Li$^*$,
	Boxun Fu,
	Zhifu Zhao
	and Xiaotian Wang
	\thanks{Tianzhi Feng, Chennan Wu, Yi Niu, Fu Li, Yang Li, Boxun Fu, Zhifu Zhao and Xiaotian Wang are with the Key Laboratory of Intelligent Perception and Image Understanding of Ministry of Education, the School of Artificial Intelligence, Xidian University, Xi’an, 710071, China.\it{($^*$Corresponding author: Fu Li, Yang Li (E-mail: fuli@mail.xidian.edu.cn, liy@xidian.edu.cn).)} \protect}
}


\maketitle

\begin{abstract}
In recent years, numerous neuroscientific studies demonstrate that specific areas of the brain are connected to human emotional responses, with these regions exhibiting variability across individuals and emotional states. To fully leverage these neural patterns, we propose an Adaptive Progressive Attention Graph Neural Network (APAGNN), which dynamically captures the spatial relationships among brain regions during emotional processing. The APAGNN employs three specialized experts that progressively analyze brain topology. The first expert captures global brain patterns, the second focuses on region-specific features, and the third examines emotion-related channels. This hierarchical approach enables increasingly refined analysis of neural activity. Additionally, a weight generator integrates the outputs of all three experts, balancing their contributions to produce the final predictive label. Extensive experiments conducted on SEED, SEED-IV and MPED datasets indicate that our method enhances EEG emotion recognition performance, achieving superior results compared to baseline methods.
\end{abstract}


\IEEEpeerreviewmaketitle

\section{Introduction}
Emotions significantly impact various aspects of human daily life and psychological health, influencing our decision-making, motivation, attention, memory, problem-solving abilities \cite{Gyurak2011,Scherer2019}. Consequently, it is necessary to develop methods to objectively and accurately identify human emotions. The ability for machines to understand human feelings has become an important field of study, drawing significant attention from Human–Machine Interaction (HMI) and pattern recognition researchers lately \cite{Wang2023,Pan2023,Liu2023,Zhou2023}. Most of these studies primarily utilize two typical signals, i.e., external and internal responses. External responses mainly include some behavioral data, such as facial expression\cite{IOANNOU2005423}, speech signals\cite{kwon2003emotion}, conversational data on social media platforms\cite{ghosal2019dialoguegcn}. Internal responses are based on physiological signals, including electromyography (EMG)\cite{cheng2008emotion}, electrocardiogram (ECG)\cite{agrafioti2011ecg}, and electroencephalogram (EEG)\cite{zheng2016multichannel}. Neuroscience research suggests that physiological signals provide more direct access to emotional origins than behavioral indicators. Thus, more and more researchers focus on this field during the past several years.

EEG is a widely used technique for recording the electrophysiological activity of neurons in the cerebral cortex via electrodes attached to the scalp \cite{zhong2020eeg}. As a typical physiological signal, EEG has shown considerable promise in decoding human emotions \cite{Wu2023,Xue2023,Tian2023,Gong2023}. Over the past decades, numerous approaches have emerged for efficiently decoding emotions from EEG signals. Among various methods, convolutional neural networks (CNNs) are extensively employed in recognizing emotions from EEG data. 
For example, Hasan et al. \cite{Hasan2021} utilize Fast Fourier Transform (FFT) and CNNs to classify 64 emotions, achieving notable accuracy in valence and arousal dimensions. Similarly, Ali et al. \cite{Ali2020} utilize a Capsule Network (CapsNet) that outperforms traditional support vector machines (SVMs) and CNNs, achieving average accuracies of 80.22\% and 85.41\%, respectively.  Ahmad et al. \cite{Ahmad2021} focused on a CNN-based model for classifying emotions into positive, neutral, and negative categories, attaining over 93\% accuracy across all categories. 
Recent advancements include He et al.'s \cite{He2023} CNN architecture for complex EEG signal classification and Aldawsari et al.'s \cite{Aldawsari2023} optimized 1D-CNN-based process, which significantly improve efficiency. Hybrid models combining CNNs with other techniques, such as LSTM and ensemble learning, have also shown promising results. As an example, Ali et al. \cite{Ali2022} and Yuvaraj et al. \cite{Yuvaraj2023} demonstrate high accuracies, with Ali’s model reaching an impressive 98\%.  
Other notable innovations include the integration of differential entropy with CNN-BiLSTM by Cui et al. \cite{Cui2022}, which achieves over 94\% accuracy on DEAP and SEED datasets, and the introduction of CIT-EmotionNet by Lu et al. \cite{Lu2023}, which combines CNN and Transformer models to outperform previous methods. Huang et al. \cite{Huang2023} incorporate attention mechanisms within a CNN-BiLSTM framework, achieving near-perfect accuracies in multi-class tasks. Further studies, such as those by Saha et al. \cite{Saha2023}, Li et al. \cite{Li2023}, and Tao et al. \cite{Tao2023}, explore wavelet decomposition, multi-scale CNNs, and attention mechanisms, all contributing to improved emotion recognition performance from EEG signals. Wang et al. \cite{Wang2023} introduces self-supervised learning to CNNs, enhancing both resource utilization and performance. 
Recent approaches include Jin et al.'s \cite{Jin2023} CNN-Transformer network for fNIRS-based emotion recognition and Farokhah et al.'s \cite{Farokhah2023} simplified 2D CNN with selective channel to improve inter-subject accuracy. Asif et al. \cite{Asif2023} demonstrate the potential of combining various convolution layers within a CNN framework for subject-independent tasks.

While these algorithms have shown impressive performance, they often overlook the inter-channel topological relationships inherent in EEG channel spatial features. This limitation arises from CNNs' inability to process non-Euclidean spaces, such as graphs and manifolds. To better capture the spatial relationships within EEG signals, several Graph Neural Network (GNN)-based approaches have been proposed, yielding promising results. Zheng et al. \cite{Zheng2021} propose a Hierarchy Graph Convolution Network (ERHGCN) that achieves classification performance of 90.56\% for valence dimension and 88.79\% for arousal dimension, demonstrating the potential of hierarchical GCN structures for emotion recognition. Saboksayr et al. \cite{Saboksayr2021} apply Graph Signal Processing (GSP) techniques, showing enhanced performance when compared with traditional methods and highlighting the efficacy of graph-based approaches. Gilakjani et al. \cite{Gilakjani2023} combine GNNs with contrastive learning and GAN-based data augmentation, leading to enhanced classification performance on both DEAP and MAHNOB-HCI datasets. Li et al. \cite{Li2023} develop a model that learns discriminative graph topologies in EEG networks, achieving an average accuracy of 84.56\% in online experiments, which is particularly important for affective brain-computer interfaces. Klepl et al. \cite{Klepl2024} present a comprehensive survey on GNN applications in EEG signal classification, providing valuable insights into their advantages and limitations.
In the realm of deep learning, Abdulrahman et al. \cite{Abdulrahman2022} achieve notable accuracies of 70.89\% in binary classification and 90.33\% for multi-class emotion recognition tasks using advanced deep learning models. DGCNN, proposed by Song et al. \cite{song2018eeg}, dynamically learns the graph's adjacency matrix to establish spatial relationships. Zhong et al.'s RGNN \cite{zhong2020eeg} enhances GNN model robustness against cross-subject variations and noisy labels by incorporating two regularization techniques. The IAG model by Song et al. \cite{song2020instance} adaptively generates directed graph connections from input graphs, allowing for exploration of intrinsic relationships between EEG regions.
 
Although considerable progress has been made in emotion recognition using EEG, current approaches still face challenges in dynamically capturing the complex relationships between emotional patterns and brain functional regions. One major challenge is the significant variation in emotion-related brain activity between individuals, which demands more adaptive recognition approaches. To address these challenges, we propose a novel Adaptive Progressive Attention Graph Neural Network (APAGNN) that finely screens out critical EEG channels across different subjects through three specialized experts. Specifically, three experts work collaboratively through a progressive analysis pipeline. The first expert analyzes global brain topology patterns, followed by the second expert that identifies emotion-specific regions, while the third expert focuses on critical channels. This hierarchical refinement process enables a comprehensive understanding of emotion-related neural patterns at multiple levels of granularity. The APAGNN model incorporates several innovative components to enhance its effectiveness. To encourage diversity in feature learning while preventing redundancy among experts, we develop a diversity-preserving training strategy that maximizes Jensen-Shannon (JS) Divergence among expert probability distributions. Additionally, we design a dynamic expert fusion method that optimally integrates the outputs from multiple experts for final classification.

To our knowledge, this is the first work to exploit the discrimination for emotional EEG expression from global to region brain. Our experimental results highlight the effectiveness of this progressive attention model. Besides, we also investigate the impact that varying numbers of experts have on the performance of emotion recognition, compare dynamic and static attention mechanisms, and analyze how these factors influence the overall accuracy of emotion classification.

\begin{figure*}[htbp]
    \centering
    \includegraphics[width=1\textwidth]{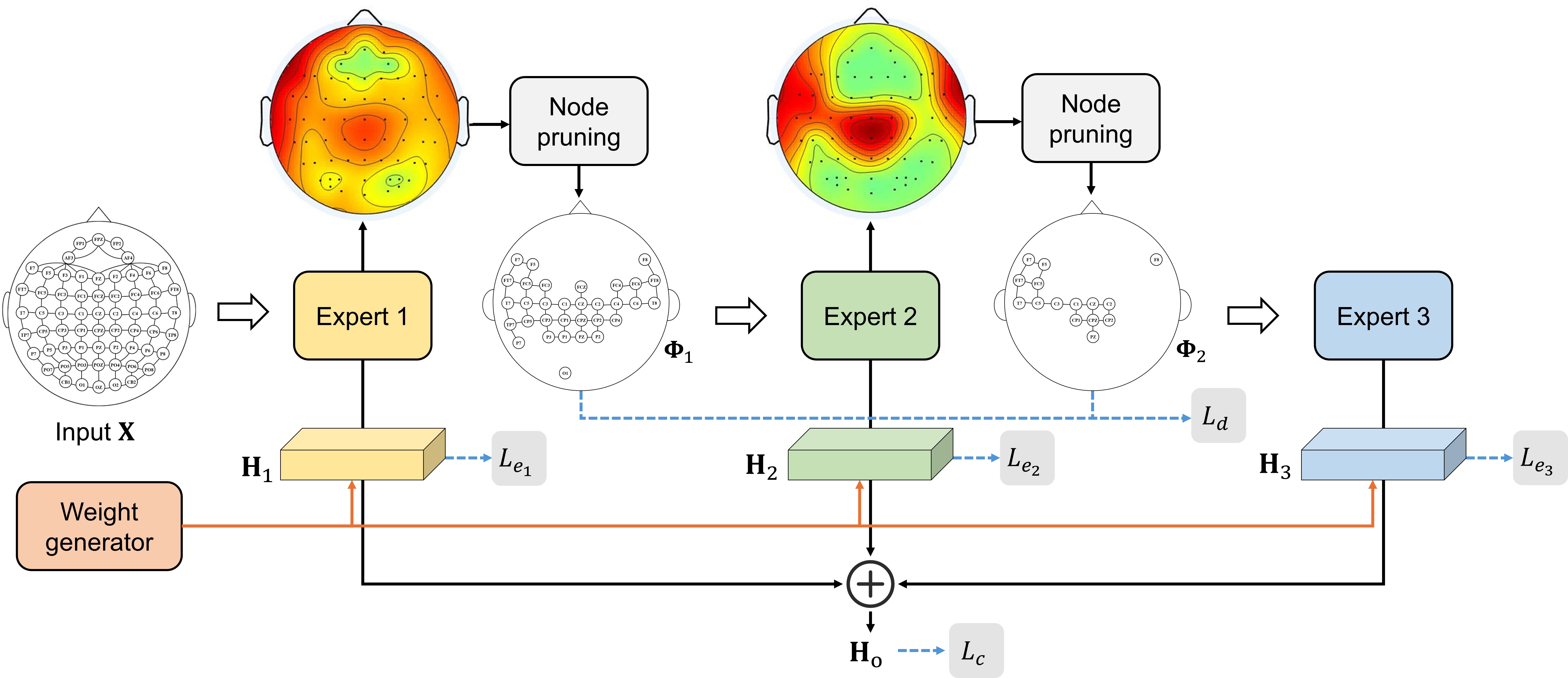}
    \caption{The architecture of APAGNN. In the multi-expert architecture, each expert learns discriminative features for emotion classification while generating attention maps to guide subsequent experts. The first two experts progressively refine the graph structure through attention operations and node pruning, optimizing the input for the next expert in the sequence. The dynamic expert fusion module then integrates the experts' predictions using a weight generator to produce the final model output.}
    \label{fig:overall_architecture}
\end{figure*}

\section{Preliminaries}
In this section, we provide the theoretical preliminaries on GNNs and attention mechanisms, which are the basis of our proposed APAGNN method.
 
\subsection{Graph Neural Network}
We represent an undirected and connected graph as $\mathcal{G}=(\mathcal{V},\mathcal{E})$, where $\mathcal{V}$ is the set of vertices (or nodes) and $\mathcal{E}$ corresponds to the set of edges, which are unordered pairs of vertices. For a graph with $|\mathcal{V}|=N$ nodes, its structure can be characterized by the adjacency matrix $\mathbf{A}\in\mathbb{R}^{N\times N}$, where $\mathbf{A}_{ij}=1$ if a connection exists between vertices $i$ and $j$, and $\mathbf{A}_{ij}=0$ otherwise.

The degree matrix $\mathbf{D}\in\mathbb{R}^{N \times N}$ is defined as a diagonal matrix where each diagonal element $\mathbf{D}_{ii}$ represents the degree of vertex $i$, computed as $\mathbf{D}_{ii}=\sum_{j=1}^N \mathbf{A}_{ij}$. From these matrices, we can derive the normalized Laplacian matrix $\mathbf{L}=\mathbf{I}_N-\mathbf{D}^{-1/2}\mathbf{A}\mathbf{D}^{-1/2}\in\mathbb{R}^{N\times N}$, where $\mathbf{I}_N$ is the $N\times N$ identity matrix. The Laplacian matrix plays a fundamental role in graph signal processing, as it encodes the graph structure and admits an eigendecomposition $\mathbf{L}=\mathbf{U}\mathbf{\Lambda}\mathbf{U}^\top$, where $\mathbf{U}$ contains the eigenvectors and $\mathbf{\Lambda}$ contains the corresponding eigenvalues.

In graph convolutional networks (GCNs), signals on the graph $\mathbf{x}\in\mathbb{R}^N$ can be transformed from the spatial domain to the spectral domain using the graph Fourier transform (GFT), defined as $\hat{\mathbf{x}}=\mathbf{U}^\top\mathbf{x}$. The convolution operation on the graph is then formulated in the spectral domain as a filtering operation  $\mathbf{y}=\mathbf{U} g_{\theta}(\mathbf{\Lambda})\mathbf{U}^\top\mathbf{x}$, where $g_\theta(\mathbf{\Lambda})$ is a learnable filter in the spectral domain that applies spectral weighting based on the eigenvalues in $\mathbf{\Lambda}$.

To address the challenges of localized filtering and computational complexity, Defferrard et al. \cite{defferrard2016convolutional} introduced Chebyshev polynomials to approximate the spectral filters. This approach leverages the properties of Chebyshev polynomials to construct an efficient filtering process by considering only a limited range of eigenvalues. Consequently, the convolution operation becomes more efficient, enabling the practical application of GCNs to larger graphs while maintaining the ability to capture localized characteristics.

\subsection{Attention Mechanism}
The concept of attention in deep learning was inspired by human visual attention behaviors. During visual processing, humans concentrate on salient regions that are pertinent for decision-making while disregarding less relevant portions \cite{xu2015show}. Similarly, in deep learning applications, certain parts of the input may hold more significance for decision-making than others. Over the past few years, various attention-based approaches have been designed to capture this selective processing, including Class Activation Mapping (CAM), Gradient-weighted Class Activation Mapping (Grad-CAM), Saliency Maps, and both hard and soft attention strategies. The integration of these attention modules into neural networks has demonstrated significant performance improvements across numerous studies \cite{xu2021grad,wang2019sharpen,gao2021ts,zhang2019learning}.

In computer vision tasks, feature maps from the final convolutional layer are known to encode rich semantic information. These feature maps are weighted, summed, and then upscaled to the original image size to obtain the CAM for a specific category. The resulting attention map highlights regions of the input that are most influential on classification decisions. Grad-CAM, proposed by Selvaraju et al. \cite{selvaraju2017grad}, extends the original CAM approach by leveraging class-specific gradient information as weights for the feature maps. Thus, Grad-CAM works with many different types of network structures without the need for architectural modifications. The Grad-CAM computation is formally expressed as $L_{\text{Grad-CAM}}^{c} = \operatorname{ReLU}\left( \sum_{k} \alpha_{k}^{c} F_{k} \right)$, where $L_{\text{Grad-CAM}}^{c}$ represents the corresponding Grad-CAM for category $c$, and $F_k$ denotes the $k$-th feature map from the final convolutional layer. The ReLU activation ensures that only features positively contributing to the class prediction are preserved in the attention map. The weight coefficient $\alpha_{k}^{c}$ for the $k$-th feature map is calculated by backpropagating the gradients of the specific class score $y^c$ with respect to the feature map $F_k$: $\alpha_{k}^{c} = \frac{1}{WH} \sum_{i=0}^{W} \sum_{j=0}^{H} \frac{\partial y^c}{\partial F_{k}(i,j)}$, where $W$ and $H$ are the width and height of feature map $F_k$, respectively, and $F_{k}(i,j)$ represents the activation value of the $k$-th feature map at spatial location $(i,j)$. The score $y^c$ corresponds to the logit or probability for class $c$ from the network's output. The weighting coefficient $\alpha_k^c$ signifies the importance of the $k$-th feature map $F_k$ concerning the target class $c$, effectively measuring its contribution to the prediction score $y^c$.

\begin{figure*}[htbp]
    \centering
    \includegraphics[width=1\textwidth]{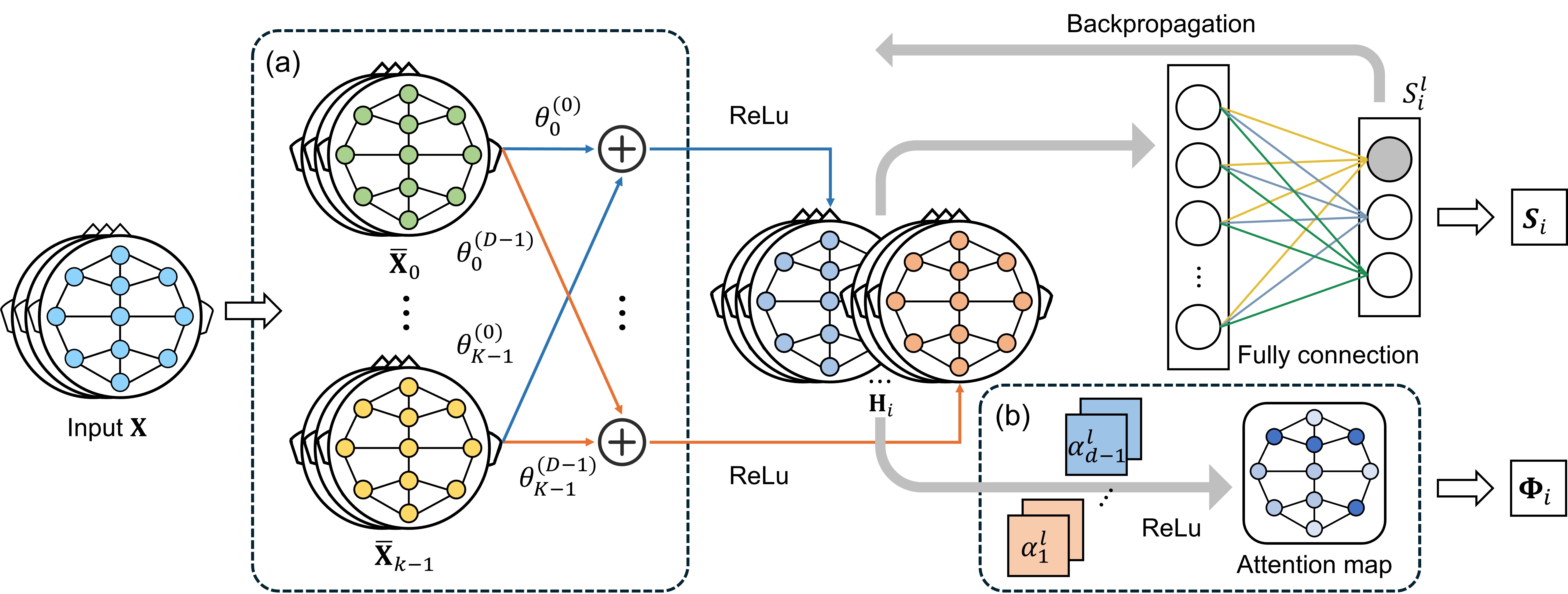}
    \caption{Schematic illustration of the expert module. Each expert performs two parallel tasks: (a) graph convolution operations for feature learning and (b) attention map generation for channel importance weighting. The module produces dual outputs: classification decisions and corresponding attention maps that highlight the influential EEG channels contributing to these decisions.}
    \label{fig: expert}
\end{figure*}

\section{Methodology}
For clarity regarding our methodology, Fig. \ref{fig:overall_architecture} presents the APAGNN model structure. Its goal is to capture more discriminative EEG representation for emotion recognition. We employ three steps to attain this objective. The first step focuses on building the spatial relationships among brain regions from global to local regions progressively. Subsequently, to encourage diversity in feature learning while preventing redundancy among experts, we implement a knowledge diversity-preserving operation. Third, a dynamic expert fusion strategy is designed to integrate the knowledge from multiple experts for final emotion prediction. We detail the complete procedure below.

\subsection{Multi-Expert Progressive Learning}
To model the complex spatial relationships among different neural regions at multiple scales, we employ three specialized experts that progressively analyze brain activity from global patterns to regional features and ultimately to electrode-specific characteristics. Each expert focuses on a different spatial scale, allowing for both broad and fine-grained feature extraction. The structure of each expert is presented in Fig. \ref{fig: expert}. Each expert performs two primary tasks: learning discriminative features for emotion classification and generating attention maps to transfer knowledge. By utilizing three experts with this structure, we achieve the goal of progressive learning.

\textbf{Graph-based EEG representation Learning.} 
To effectively capture the spatial dependencies in EEG signals, we leverage GNNs with Chebyshev filters to construct each expert. 
The process begins by transforming raw EEG samples into graph representations, where each EEG electrode serves as a vertex. Specifically, we first extract differential entropy (DE) features from five frequency bands. These features serve as the initial node attributes, yielding a feature matrix $\mathbf{X}\in\mathbb{R}^{C\times F}$, in which $C$ represents the electrode count, and $F$ indicates the frequency band count. To model the spatial relationships between electrodes, we construct an adjacency matrix $\mathbf{A} \in \mathbb{R}^{C \times C}$ based on the physical arrangement of the electrodes \cite{li2022gmss}. Through this process, each raw EEG sample is converted into a graph structure $\mathcal{G} = (\mathcal{V}, \mathcal{E}) = (\mathbf{X}, \mathbf{A})$, with $\mathcal{G}$ denoting a graph with vertices $\mathcal{V}$
and edges $\mathcal{E}$.
The graph convolution operation, implemented using Chebyshev polynomial formulation, is given by:
\begin{equation}
	\label{graph convolution_approach_1}
	\mathbf{y} = \sum_{k=0}^{K-1}\theta_{k}T_{k}(\widetilde{\mathbf{L}})\,\mathbf{X},
\end{equation}
where $\mathbf{\theta}_{k}$ represents learnable parameters for the $k$-th order Chebyshev polynomial, and $T_{k}(\widetilde{\mathbf{L}})$ denotes the $k$-th order Chebyshev polynomial evaluated at the scaled Laplacian $\widetilde{\mathbf{L}}$. To efficiently compute higher-order Chebyshev polynomials, we use the recurrence relation. Define $\bar{\mathbf{X}}_k = T_k(\widetilde{\mathbf{L}})\mathbf{X}$, with $\bar{\mathbf{X}}_0 = \mathbf{X}$, $\bar{\mathbf{X}}_1 = \tilde{\mathbf{L}}\mathbf{X}$, and $\bar{\mathbf{X}}_k = 2\widetilde{\mathbf{L}}\bar{\mathbf{X}}_{k-1} - \bar{\mathbf{X}}_{k-2}$. Thus, we can formulate the convolution operation on graphs as:
\begin{equation}
	\mathbf{y}
	= \theta_0\bar{\mathbf{X}}_0 + \theta_1\bar{\mathbf{X}}_1 + ... + \theta_{K-1}\bar{\mathbf{X}}_{K-1}.
\end{equation}

For a single Chebyshev filter, the learnable coefficients are represented by $\Theta_i=[\theta_0, \theta_1, ..., \theta_{K-1}]^T\in\mathbb{R}^{K\times 1}$. To learn diverse feature transformations, we implement $D$ parallel filters, each learning a different transformation of the node features. The complete set of learnable parameters is denoted as $\Theta=[\Theta_0, \Theta_1, ..., \Theta_{D-1}]\in\mathbb{R}^{K\times D}$. After computing the outputs for each node and filter across all $K$ Chebyshev orders, we concatenate the results and apply the weights $\Theta$. Finally, through the ReLU activation function, the learned EEG representation $\mathbf{H}_i\in\mathbb{R}^{C\times D}$ for the $i$-th expert is obtained:
\begin{equation}
	\mathbf{H}_i = \operatorname{ReLU}([\bar{\mathbf{X}}_0, \bar{\mathbf{X}}_1, ..., \bar{\mathbf{X}}_{K-1}]\times\Theta).
\end{equation}
    
The learned features $\mathbf{H}_{i}$ are then flattened to input to a fully connected layer to obtain class probability scores $\mathbf{S}_{i} \in\mathbb{R}^{E}$ for each emotion class, with $E$ indicating the total emotion categories. The $i$-th expert's classification loss is calculated using:
\begin{equation}
L_{e_{i}} = -\sum_{l=1}^{E} y_i^l \log(S_{i}^{l}),
\end{equation}
where $y_i^l$ represents the one-hot encoded actual label for class $l$, and $S_i^l$ represents the probability of class $l$ predicted by the $i$-th expert.

\textbf{Attention-guided Knowledge Transfer.} 
To facilitate knowledge transfer across experts, we construct an attention map that transfers important electrodes identified by earlier experts to subsequent ones. This allows later experts to build upon patterns discovered by previous ones. We achieve this through dynamic attention method \cite{selvaraju2017grad}, which helps us identify the electrodes that contribute significantly to the final decision.

Specifically, for a target emotion class index $l$, we calculate the gradient between predicted probability $S_{i}^{l}$ and feature representations $\mathbf{H}_{i}$ through backpropagation. The gradients are processed through global average pooling to derive the importance score $\alpha_d^l$ for each feature dimension $d$:
\begin{equation}
    \alpha_d^l = \frac{1}{C} \sum_{c=0}^{C-1} \frac{\partial S_i^{l}}{\partial \mathbf{H}_i},
\end{equation}
This gradient-based approach reflects the significance of each feature dimension on the classification decision.
Next, the weights $\alpha_d^l$ and the feature map $\mathbf{H}_{i}$ are linearly combined to generate the attention map for class $l$. By taking the mean of attention weights $\alpha_d^l$  across feature dimensions, we derive the final channel importance $\mathbf{I}_{i}$:
\begin{equation}
        \mathbf{I}_{i} = \operatorname{ReLU}\left(\sum_{d=0}^{D-1} \alpha_d^l \mathbf{H}_i \right),
\end{equation}
where $\operatorname{ReLU}$ ensures only positive contributions are preserved. 

Finally, using the attention map $\mathbf{I}_i$ from each expert, we selectively prune weakly emotion-related connections. To ensure consistent thresholding across different attention scales, we first normalize $\mathbf{I}_i$ to the range $[0,1]$:
\begin{equation}
\widetilde{\mathbf{I}}_i 
= \frac{\mathbf{I}_{i} - \operatorname{min}(\mathbf{I}_{i})}{\operatorname{max}(\mathbf{I}_{i}) - \operatorname{min}(\mathbf{I}_{i})}.
\end{equation}
Based on this normalized attention map, we establish a threshold $\eta \in (0,1)$. Any channel-node with an activation value below this threshold is considered for pruning. To maintain data structure integrity during training, the pruned node's value is set to zero, and its associated edges are removed by zeroing out the relevant rows and columns in the adjacency matrix. This process yields the attention map $\mathbf{\Phi}_i$, highlighting the important electrodes learned by the $i$-th expert, which will be transferred to the next expert.

\textbf{Expert Progressive Learning.} 
The three experts work in sequence to progressively refine the analysis of EEG signals. The pipeline begins with the first expert conducting a global-scale analysis on the original graph $\mathcal{G}$, generating both feature representation $\mathbf{H}_{1}$ and attention map $\mathbf{\Phi}_{1}$ that identifies emotionally salient regions. Building upon this foundation, the second expert receives both the original graph $\mathcal{G}$ and the attention map $\mathbf{\Phi}_1$ from the first expert to perform region-scale analysis. By utilizing $\mathbf{\Phi}_1$ to mask irrelevant nodes in $\mathcal{G}$, it focuses on the emotion-relevant areas. The second expert then applies the same graph convolution and attention computation process with its own learnable parameters, producing a refined feature representation $\mathbf{H}_2$ and an attention map $\mathbf{\Phi}_2$ with important electrodes. The final stage of analysis is performed by the third expert, which combines $\mathcal{G}$ and $\mathbf{\Phi}_2$ to conduct electrode-scale analysis, yielding the feature representation $\mathbf{H}_3$ that captures fine-grained spatial patterns within EEG signals. 

\subsection{Knowledge Diversity Preserving}
To encourage knowledge diversity among different experts and reduce redundancy in their focus areas, we introduce a diversity-preserving training method based on Jensen-Shannon (JS) Divergence. This approach aims to maximize the JS divergence between attention maps produced by the first two experts, thereby promoting the extraction of distinct EEG emotion-related attention patterns. 

Specifically, we normalize the attention maps $\boldsymbol{\Phi}_1$ and $\boldsymbol{\Phi}_2$ from the first and second experts to ensure they represent valid probability distributions. This is achieved by applying the softmax function across the EEG channels for each attention map:
\begin{equation}
\boldsymbol{\Phi}_{1}^{'}=\frac{\exp \left(\boldsymbol{\Phi}_{1}^c\right)}{\sum_{c=1}^{C} \exp \left(\boldsymbol{\Phi}_{1}^c\right)}, 
\end{equation}
\begin{equation}
\boldsymbol{\Phi}_{2}^{'}=\frac{\exp \left(\boldsymbol{\Phi}_{2}^c \right)}{\sum_{c=1}^{C} \exp \left(\boldsymbol{\Phi}_{2}^c \right)},
\end{equation}
where $C$  represents the total number of EEG electrodes. The JS divergence between the two normalized attention distributions $\boldsymbol{\Phi}_{1}^{'}$ and $\boldsymbol{\Phi}_{2}^{'}$ from the first and second experts is defined as:
\begin{equation}
	\label{JS-formula}
	\begin{aligned}
		L_d = \operatorname{JS} (\boldsymbol{\Phi}_{1}^{'}\,||\,\boldsymbol{\Phi}_{2}^{'}) = & \frac{1}{2} \operatorname{KL} \left( \boldsymbol{\Phi}_{1}^{'}\,||\frac{\boldsymbol{\Phi}_{1}^{'}+\boldsymbol{\Phi}_{2}^{'}}{2} \right) \,+ \\
		& \frac{1}{2} \operatorname{KL} \left( \boldsymbol{\Phi}_{2}^{'}\,||\frac{\boldsymbol{\Phi}_{1}^{'}+\boldsymbol{\Phi}_{2}^{'}}{2} \right),
	\end{aligned}
\end{equation}
\noindent where $\operatorname{KL}(\cdot \| \cdot)$ denotes the Kullback-Leibler (KL) Divergence.

\subsection{Dynamic Expert Fusion}
To fully leverage the emotion-related features extracted by the three experts, we integrate their representations through a dynamic weight generator. The weight generator assigns importance coefficients $\xi_1, \xi_2, \xi_3$ to each expert's representation $\mathbf{H}_1$, $\mathbf{H}_2$, and $\mathbf{H}_3$, yielding the final output:
\begin{equation}
	\label{My-MoE}
	\mathbf{H}_\text{o} = \sum_{i=1}^{3} \xi_i \cdot \mathbf{H}_i.
\end{equation}
Then the final data representation $\mathbf{H}_\text{o}$ is flattened and passed to a fully-connected layer before applying a softmax transformation to generate the predicted probability distribution $P(l\,|\,\mathbf{X})$, which indicates the probability that the EEG sample $\mathbf{X}$ is classified as the $l$-th emotional state:
\begin{equation}
P(l \mid \mathbf{X}) = \frac{\exp(z^l)}{\sum_{e=0}^{E-1} \exp(z^{e})},
\end{equation}
where $E$ is the total number of emotion categories, $z^l$ represents the logit (pre-softmax activation) for the $l$-th emotion category. 

Finally, the cross-entropy loss, which quantifies the discrepancy between predicted probability outputs and ground-truth label, is defined as:
\begin{equation}
	\label{Loss_1}
	L_{c} = - \log P\left(l_\text{gt} \mid \mathbf{X} \right),
\end{equation} 
where $l_\text{gt}$ represents the ground-truth label for the EEG sample $\mathbf{X}$.

Finally, we compute the total loss $L_\text{total}$ by averaging the combination of prediction loss, expert classification losses, and diversity loss across all training samples in a mini-batch:
\begin{equation}
    \label{loss}
    L_\text{total} =\frac{1}{N}\sum_{n=1}^{N} \left(L_{c}(n)+ \lambda \cdot \sum_{i=1}^{3} L_{e_{i}}(n)+\beta \cdot L_{d}(n)\right),
\end{equation}
where $N$ represents the sample count within each mini-batch, and $\lambda$ and $\beta$ function as hyperparameters that control the contributions of experts' losses and the JS divergence term, respectively.

\section{Experiment}

\subsection{Datasets}
To validate the proposed APAGNN model, we conducte extensive experiments on three open-access databases, which record EEG data while subjects are viewing emotion-eliciting video clips.

$\mathbf{SEED}$: This dataset includes EEG signals from 15 participants, each participating in three sessions. During each session, the subjects watch 15 video clips, categorized into three emotional states: happy, neutral, and sad. The raw EEG signals are sampled every second and passed through a 0.3–50 Hz bandpass filter. For experimental consistency, we followed the same protocol as \cite{zheng2015investigating}, using data from the initial nine trials for training and the final six trials for testing.

$\mathbf{SEED}$-$\mathbf{IV}$: This dataset resembles SEED but includes four emotion types, with a total of 24 video clips per session (6 clips per emotion). We adhered to the experimental protocol in \cite{zheng2018emotionmeter}, where data from the first sixteen trials is used for training and the final eight trials are utilized for testing.

$\mathbf{MPED}$: This dataset comprises EEG signals from 30 subjects (one session per subject) and encompasses seven emotion types. For each emotion category, participants viewed four different video clips, yielding 28 trials per session. Following the protocol in \cite{song2019mped}, we first obtain Short-Time Fourier Transform (STFT) features from five frequency bands, and then use 21 trials for model training and reserve the final seven trials for testing purposes.

\subsection{Implementation Details}
For our experiment, the node-pruning threshold $\eta$ is set to 0.5. The Chebyshev kernel size $K$ is set to 3, and the number of convolution filters $D$ is set to 32. Optimization is performed using the Adam optimizer. We train the model for 100 epochs using a batch size of 64 and a learning rate of 1e-3. Model performance is evaluated using mean accuracy (ACC) and standard deviation (STD). The model is trained with PyTorch on a GeForce RTX 2080Ti GPU.

\subsection{Experiment Results}

The first category includes GNN-based methods such as DGCNN \cite{song2018eeg}, RGNN \cite{zhong2020eeg}, IAG \cite{song2020instance}, and V-IAG \cite{9373917}. These methods use graph neural networks to analyze EEG data.
The second category consists of approaches that focus on specific brain regions or important electrode channels. This group includes BiDANN \cite{li2018novel}, BiHDM \cite{li2020novel}, and DBN \cite{zheng2015investigating}.
Additionally, we compare our model with traditional machine learning methods, specifically SVM \cite{suykens1999least} and standard GNN \cite{defferrard2016convolutional}, which are common baseline techniques in this field.

\begin{table}[h]
    \caption{Comparison of accuracies and standard deviations (\%) for the subject-dependent experiments on SEED, SEED-IV, and MPED datasets.}
    \label{table:subject-dependent_results}
    \centering
    \begin{tabular}{lccc}
        \toprule
        \textbf{Method} & \multicolumn{3}{c}{\textbf{ACC / STD (\%)}} \\
        \cmidrule(lr){2-4}
        & SEED & SEED-IV & MPED \\
        \midrule
        SVM     & 83.99 $\pm$ 09.72 & 56.61 $\pm$ 20.05 & 32.39 $\pm$ 09.53 \\
        GNN     & 87.40 $\pm$ 09.20 & 68.34 $\pm$ 15.42 & 33.26 $\pm$ 06.44 \\
        DBN     & 86.08 $\pm$ 08.34 & 66.77 $\pm$ 07.38 & 35.07 $\pm$ 11.25 \\
        BiDANN  & 92.38 $\pm$ 07.04 & 70.29 $\pm$ 12.63 & 37.71 $\pm$ 06.04 \\
        BiHDM   & 93.12 $\pm$ 06.06 & 74.35 $\pm$ 14.09 & 40.34 $\pm$ 07.59 \\
        DGCNN   & 90.40 $\pm$ 08.49 & 69.88 $\pm$ 16.29 & 32.37 $\pm$ 06.08 \\
        RGNN    & 94.24 $\pm$ 05.95 & 79.37 $\pm$ 10.54 & - \\
        IAG     & 95.44 $\pm$ 05.48 & -            & 40.38 $\pm$ 08.75 \\
        V-IAG   & 95.64 $\pm$ 05.08 & -            & 40.40 $\pm$ 09.35 \\
        APAGNN & \textbf{96.38 $\pm$ 04.19} & \textbf{86.64 $\pm$ 10.43} & \textbf{41.58 $\pm$ 04.61} \\
        \bottomrule
    \end{tabular}
\end{table}

As shown in Table \ref{table:subject-dependent_results}, the APAGNN model achieves superior performance compared to the existing methods on all three datasets. Specifically, it achieves accuracies of 96.38\% in SEED, 86.64\% in SEED-IV, and 41.58\% in MPED. Notably, the APAGNN model surpasses the previous best method, IAG, by 0.94\% and 1.2\% in accuracy, while reducing the standard deviations by 1.29\% and 4.14\% on the SEED and MPED datasets, respectively. Remarkably, APAGNN consistently achieves the lowest standard deviation across all three datasets, demonstrating its superior stability and robustness.

\begin{table}[!h]
    \caption{Comparison of accuracies and standard deviations (\%) with varying numbers of experts. For better comparison, here we denote our APAGNN as APAGNN-3E.}
    \label{table:expert_num}
    \centering
    \renewcommand\arraystretch{1.3}{
        \setlength{\tabcolsep}{3mm}{
            \begin{tabular}{lccc}
                \toprule
                \textbf{Method} & \multicolumn{3}{c}{\textbf{ACC ± STD (\%)}} \\
                \cmidrule(lr){2-4}
                & SEED & SEED-IV & MPED \\
                \midrule
                APAGNN-2E & 93.73 ± 05.78 & 83.59 ± 12.29 & 39.04 ± 04.28 \\
                APAGNN-3E & \textbf{96.38 ± 04.19} & \textbf{86.64 ± 10.43} & \textbf{41.58 ± 04.61} \\
                \bottomrule
            \end{tabular}
        }
    }
\end{table}

\begin{figure}[htbp]
    \centering
    \includegraphics[width=1\columnwidth]{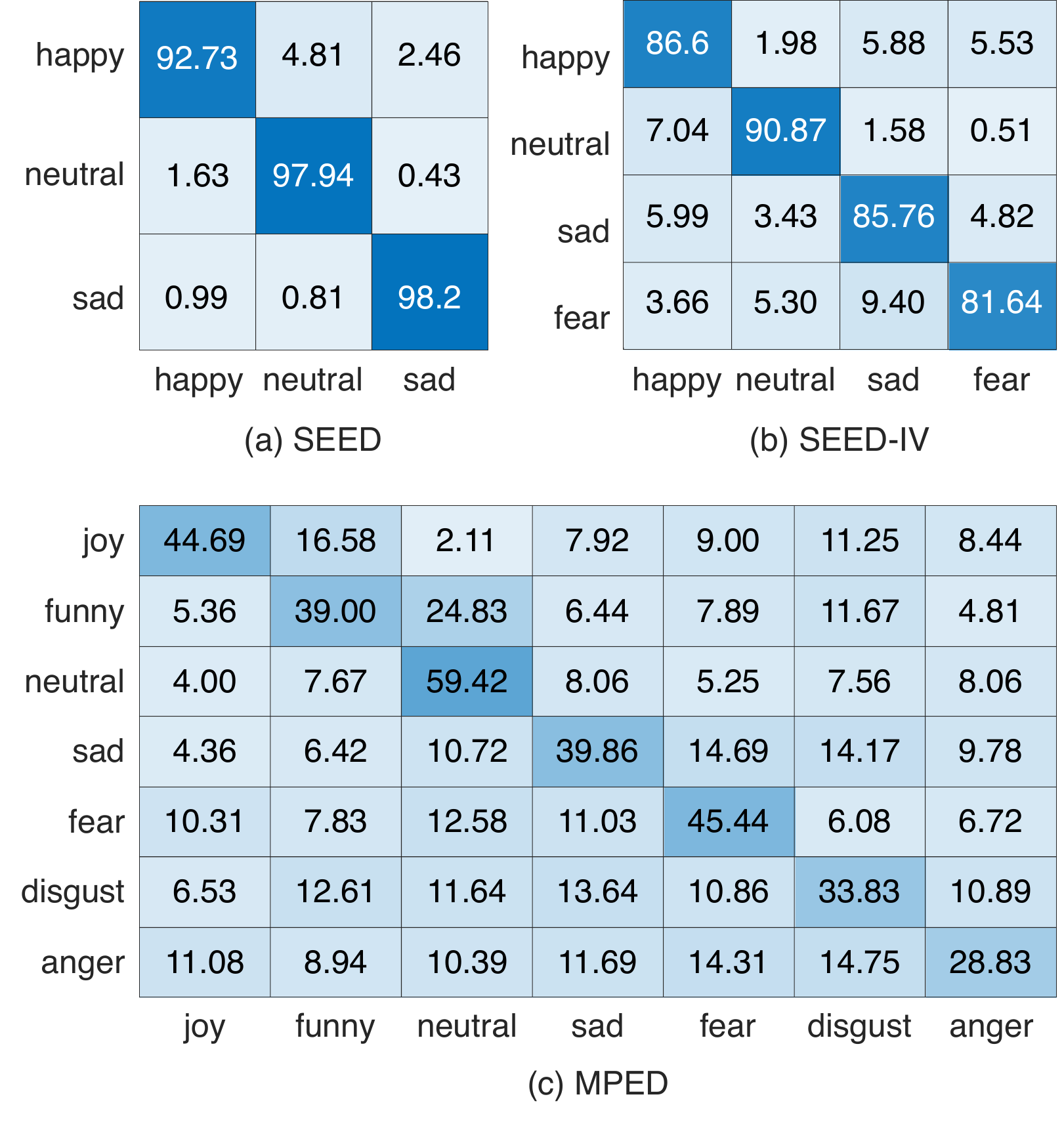}
    \caption{Confusion matrices for subject-dependent experiments on SEED, SEED-IV and MPED datasets.}
    \label{fig: confusion_matrix}
\end{figure}

\begin{figure*}[htbp]
    \centering
    \includegraphics[width=1\textwidth]{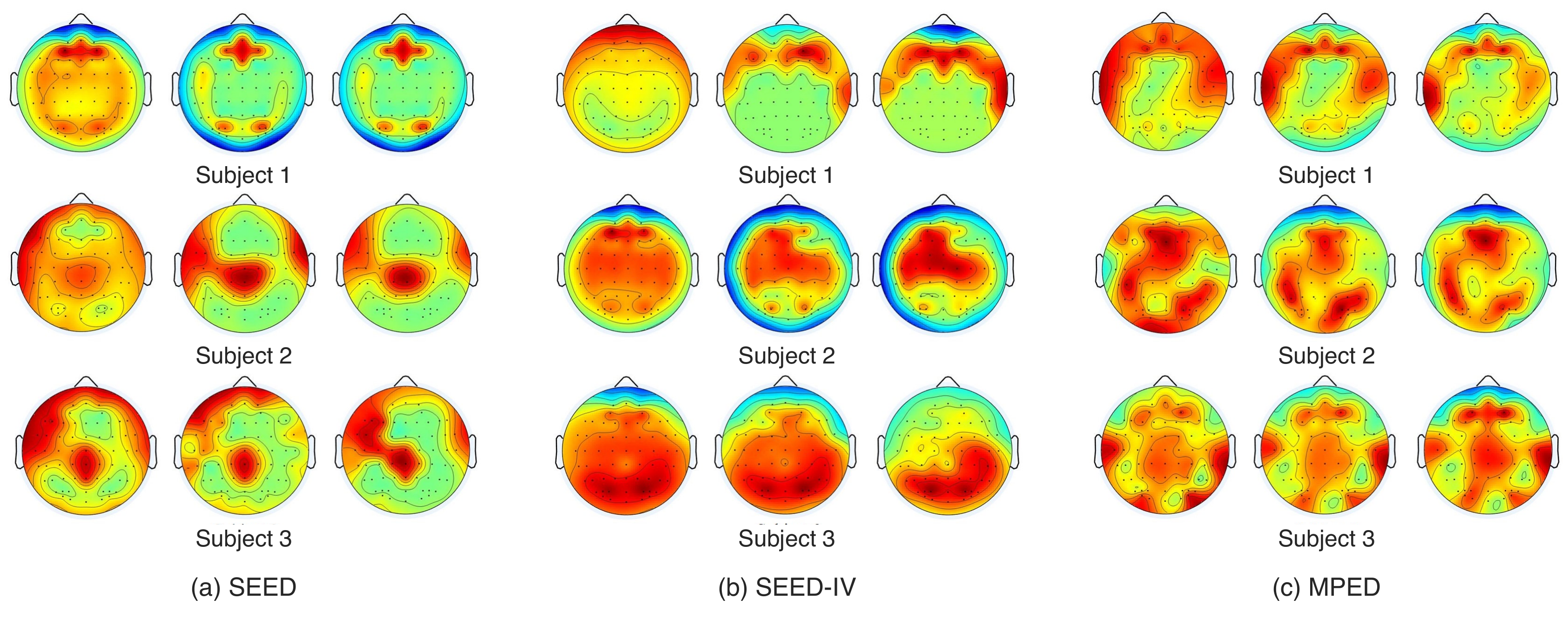}
    \caption{Attention maps generated by three experts across different subjects from three datasets: (a) SEED, (b) SEED-IV, and (c) MPED. Each row represents a different subject, while columns show the attention patterns generated by each expert in sequence. For completeness of analysis, we visualize patterns from three experts, although only the first two are implemented in the APAGNN.}
    \label{fig: attention_map}
\end{figure*}

To further confirm the progressive attention module's effectiveness, we conduct an extra experiment  by replacing the progressive learning process with single attention. The resulting model can be denoted as APAGNN-2E. For better comparison, we denote our APAGNN as APAGNN-3E here. Table \ref{table:expert_num} displays the experimental results. It is clear to see that APAGNN-3E that contains progressive attention outperforms APAGNN-2E that consists of only one attention process. The accuracy improvements are 2.65\% for SEED, 3.05\% for SEED-IV, and 2.54\% for MPED. This verifies the important effect of the proposed progressive attention. Meanwhile, we can find that the sub-graph structure, generated by the deeper experts, could further improve the performance.

\begin{table}[!t]
    \caption{Comparison of the impact of removing vs. retaining the diversity-preserving training strategy on accuracy and standard deviation (\%).}
    \label{table:constraint_effect}
    \centering
    \renewcommand\arraystretch{1.3}{
        \setlength{\tabcolsep}{3mm}{
            \begin{tabular}{lcc}
                \toprule
                \textbf{Method} & \multicolumn{2}{c}{\textbf{ACC ± STD (\%)}} \\
                \cmidrule(lr){2-3}
                & SEED & SEED-IV \\
                \midrule
                APAGNN($- L_d$) & 94.22 ± 05.76 & 84.86 ± 10.69 \\
                APAGNN & \textbf{96.38 ± 04.19} & \textbf{86.64 ± 10.43} \\
                \bottomrule
            \end{tabular}
        }
    }
\end{table}

We also conduct an ablation experiment to examine the effect that expert diversity loss $L_d$ has on model performance. The results in Table \ref{table:constraint_effect} highlight the essential role of this loss in improving classification accuracy. Specifically, when the diversity loss $L_d$ is removed from the training objective, we observe a notable decline in performance across datasets. The classification accuracy decreases by 2.16\% for SEED and 1.78\% for SEED-IV datasets, respectively. Without this strategy to maintain diversity among experts, the experts tend to focus on similar feature extraction patterns, which limits their ability to capture different aspects of EEG emotion patterns.

\begin{figure*}[htbp]
    \centering
    \includegraphics[width=0.8\textwidth]{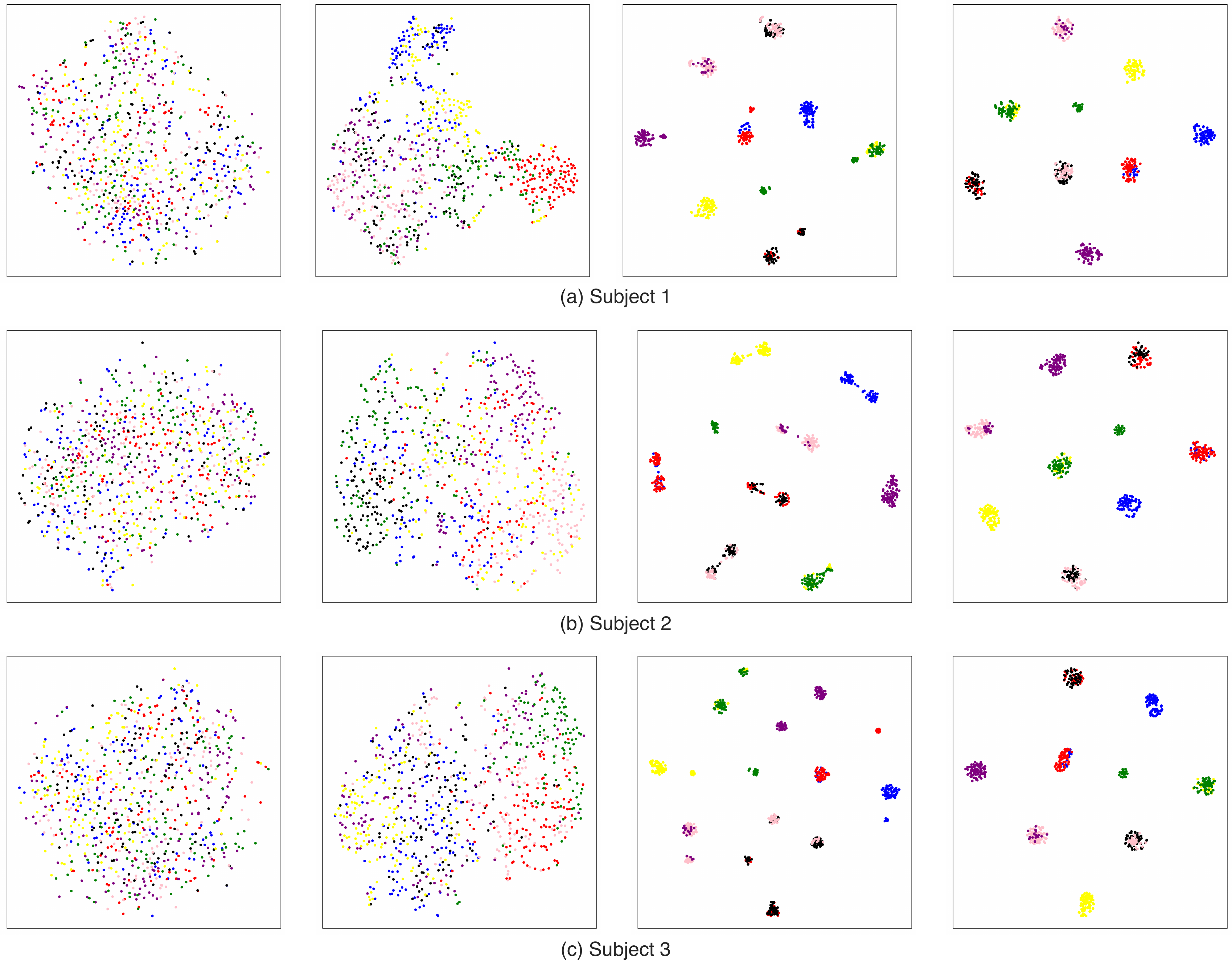}
    \caption{T-SNE visualization of the feature learning process on the MPED dataset. Each row shows the feature evolution for an individual subject across four stages: initial normalized EEG data (first column) and subsequent transformations by each expert (columns 2-4). Seven emotions are represented by distinct colors: joy (red), funny (blue), neutral (yellow), sad (pink), fear (purple), disgust (black), and angry (green).}
    \label{fig: t-sne}
\end{figure*}

\subsection{Confusion Matrix}
To better understand the performance of APAGNN, Fig. \ref{fig: confusion_matrix} displays the confusion matrices across SEED, SEED-IV, and MPED datasets. In these matrices, the columns show the predicted categories, while the rows represent the true categories. Analyzing these results reveals several important insights.

1) In the SEED three-class classification task, our model performs exceptionally well, achieving over 90\% accuracy for all three emotion categories. This high accuracy suggests that the EEG patterns for each emotion are distinct and easy to differentiate.

2) In the SEED-IV four-class classification task, which includes the addition of the fear category, we observe a slight decrease in performance for the happy, neutral, and sad emotions. Among these, sad emotion is most affected. As shown in Fig. \ref{fig: confusion_matrix}(b), the highest confusion occurs between fear and sad emotions, with 9.4\% of fear instances misclassified as sad. This suggests that negative emotions like fear and sadness may have similar underlying EEG patterns.

3) In the more challenging MPED seven-class classification task, our model performs better in recognizing neutral, joy, fear, and sad emotions, but struggles more with funny, disgust, and angry emotions. Consistent with the SEED and SEED-IV tasks, the APAGNN achieves the highest accuracy in classifying neutral emotion.

Overall, our model shows strong performance across different emotion categories, especially in recognizing happy, neutral, sad, and fear emotions, with particularly good results for neutral emotions. These findings demonstrate that our APAGNN model effectively captures the key EEG features that distinguish different emotional states.

\subsection{Attention Visualization}
To validate whether our method can successfully identify important electrodes through progressive learning, we visualize the attention maps of the first three subjects from each dataset, as shown in Fig. \ref{fig: attention_map}. These attention maps highlight important parts of the brain that play a role in processing emotions, including regions within the prefrontal and temporal lobes. These critical regions are consistent with established neuroscience findings on emotion recognition \cite{coan2004frontal, davis2001amygdala, alarcao2017emotions}. Moreover, the activation in occipital lobe, which is associated with visual processing, may be related to the video-based stimulus material used in the datasets. 
Notably, the visualization outcomes confirm our APAGNN method achieves the goal of progressive attention. Taking the subjects in Fig. \ref{fig: attention_map}(a) as an example, the first expert focuses on a broad range of  emotion-related brain regions, while the second expert, building upon the knowledge from the first expert, further concentrates on more specific regions of importance.
Meanwhile, we can see subtle individual variations on different subjects. This observation demonstrates our model's ability to both learn generally important emotion-processing areas and adapt to individual-specific neural patterns.

To evaluate the advantage of our dynamic attention module in multi-expert progressive learning architecture, we modify our APAGNN framework to two variants that utilize predefined critical channel sets from previous studies \cite{zheng2015investigating,zhong2020eeg}, denoted as SPAGNN-v1 and SPAGNN-v2. For fair comparison, we carefully design the channel selection scheme according to important brain regions. Specifically, for the first expert, both variants use identical channel selections, including FPZ, FP1, FP2, AF3, AF4, F5, F6, F7, F8, FT7, FT8, C5, C6, T7, T8, CP5, CP6, TP7, TP8, P5, P7, P8, PO5, PO6, PO7, PO8, CB1, and CB2. For the second expert, the two variants utilize different channel sets based on prior research. SPAGNN-v1 employs 12 key channels identified by Zheng et al. \cite{zheng2015investigating} (FT7, FT8, C5, C6, T7, T8, CP5, CP6, TP7, TP8, P7, P8), while SPAGNN-v2 employs eight critical channels based on Zhong et al. \cite{zhong2020eeg} (FP1, FP2, AF3, AF4, F6, F8, CB2, PO8). 
We evaluate SPAGNN-v1 and SPAGNN-v2 using the SEED and SEED-IV datasets, with detailed results presented in Table \ref{table:static_results}. SPAGNN-v1 and SPAGNN-v2 with predefined electrodes achieve classification accuracies of 93.37\% and 92.38\% on SEED, respectively. And the performance reaches 82.65\% and 81.44\% on SEED-IV, respectively. While our APAGNN with dynamic attention module demonstrates superior performance, improving accuracy by 3.01\% on SEED and 3.99\% on SEED-IV compared to the best results of both variants. These results demonstrate the advantages of dynamic attention module over static approaches.

\begin{table}[!t]
    \centering
    \caption{Comparison of the effects of static versus adaptive progressive attention mechanisms on accuracy and standard deviation (\%).}
    \label{table:static_results}
    \renewcommand\arraystretch{1.3} 
    \setlength{\tabcolsep}{4mm}    
    \begin{tabular}{lcc}
        \toprule
        \multirow{2}{*}{\textbf{Method}} & \multicolumn{2}{c}{\textbf{ACC / STD (\%)}} \\
        \cmidrule(lr){2-3}
        & SEED & SEED-IV \\
        \midrule
        SPAGNN-v1 & 93.37/06.61 & 82.65/12.33 \\
        SPAGNN-v2 & 92.38/07.25 & 81.44/11.93 \\
        APAGNN    & \textbf{96.38/04.19} & \textbf{86.64/10.43} \\
        \bottomrule
    \end{tabular}
\end{table}

\subsection{T-SNE Visualization}
To further validate our proposed model's effectiveness, we employ t-SNE \cite{van2008visualizing} to visualize APAGNN's feature learning process. Fig. \ref{fig: t-sne} presents the t-SNE visualizations for the first three subjects in the MPED dataset, where distinct emotions are denoted by different colors: red (joy), blue (funny), yellow (neutral), pink (sad), purple (fear), black (disgust), and green (anger). Each row represents an individual subject's learning progress, with columns showing (from left to right): the initial post-normalization state, and the features after processing by the first, second, and third experts, respectively. The visualizations clearly show a transformation from a messy, overlapping distribution to well-separated clusters for each emotion. This progressive clustering of EEG data points, where data from the same emotion groups together while remaining distinct from other emotions, demonstrates the model's ability to learn both consistent and discriminative features effectively.

\section{Conclusion}
This study introduces APAGNN, an innovative model for decoding emotional states from EEG signals. Our approach not only achieves leading performance but also demonstrates the capability to adaptively identify subject-specific emotion-related EEG channel sets and corresponding brain region topological subgraphs. The model's adaptive nature enables the discovery of individual differences in the spatial distribution of emotion-related brain regions. Through APAGNN's interpretable architecture, we gain valuable insights into how various parts of the brain work together during emotional experiences. Our findings suggest that emotion-related neural activity patterns exhibit significant inter-subject variability rather than following a universal template, emphasizing the importance of personalized approaches in emotion recognition systems. This work contributes to both the theoretical understanding of emotion processing in different brain regions and the practical advancement of EEG-based emotion recognition technologies.

\bibliographystyle{IEEEtran}
\bibliography{ref}

\end{document}